\documentclass[reprint,amsmath,amssymb,aps,superscriptaddress]{revtex4-1}

\usepackage{graphicx}
\usepackage{bm}
\usepackage{cmath}
\usepackage{xcolor}
\begin{document}

\title{Chiral topological photonics with an embedded quantum emitter}

\author{Mahmoud~Jalali~Mehrabad}
\affiliation{%
Department of Physics and Astronomy, University of Sheffield, Sheffield S3 7RH, UK
}%
\author{Andrew~P.~Foster}%
 \email{andrew.foster@sheffield.ac.uk}
\affiliation{%
Department of Physics and Astronomy, University of Sheffield, Sheffield S3 7RH, UK
}%
\author{Ren\'e~Dost}
\affiliation{%
Department of Physics and Astronomy, University of Sheffield, Sheffield S3 7RH, UK
}%
\author{Edmund~Clarke}
\affiliation{%
EPSRC National Epitaxy Facility, University of Sheffield, Sheffield S1 4DE, UK
}%
\author{Pallavi~K.~Patil}
\affiliation{%
EPSRC National Epitaxy Facility, University of Sheffield, Sheffield S1 4DE, UK
}%
\author{A.~Mark~Fox}
\affiliation{%
Department of Physics and Astronomy, University of Sheffield, Sheffield S3 7RH, UK
}%
\author{Maurice~S.~Skolnick}
\affiliation{%
Department of Physics and Astronomy, University of Sheffield, Sheffield S3 7RH, UK
}%
\author{Luke~R.~Wilson}
\affiliation{%
Department of Physics and Astronomy, University of Sheffield, Sheffield S3 7RH, UK
}%

\date{\today}

 \begin{abstract}
Topological photonic interfaces support topologically non-trivial optical modes with helical character. When combined with an embedded quantum emitter that has a circularly polarised transition dipole moment, a chiral quantum optical interface is formed due to spin-momentum locking. Here, we experimentally realise such an interface by integrating semiconductor quantum dots into a valley-Hall topological photonic crystal waveguide. We harness the robust waveguide transport to create a ring resonator which supports helical modes. Chiral coupling of quantum dot transitions, with directional contrast as high as $75\%$, is demonstrated. The interface also supports a topologically trivial mode, comparison with which allows us to clearly demonstrate the protection afforded by topology to the non-trivial mode.
 \end{abstract}

\maketitle

\section{Introduction}

Nano-photonics concerns the confinement and manipulation of light at the nanoscale. A significant consequence of transverse optical confinement in waveguides at this scale is the presence of an elliptically polarized electric field, which carries spin angular momentum \cite{Lodahl2017}. When a quantum emitter with a circularly polarised transition dipole moment is coupled to the waveguide, a chiral quantum light-matter interface can be realised, in which the photon spin and momentum are locked and photon-emitter interactions become direction-dependent. Such an interface has numerous potential applications, ranging from single-photon routers \cite{PhysRevA.94.063817,PhysRevA.97.023821} to optical circulators \cite{Scheucher1577} and isolators \cite{PhysRevA.90.043802}. Further intriguing prospects include leveraging chirality in quantum spin networks \cite{PhysRevA.91.042116} or for entanglement generation \cite{PhysRevB.92.155304}.

The chiral quantum optical interface was first demonstrated by coupling a semiconductor quantum dot (QD) to a dielectric nanobeam waveguide \cite{Luxmoore2013a,Luxmoore2013b}. Subsequent developments extended capabilities to include atomic \cite{PhysRevLett.110.213604,Mitsch2014,Shomroni2014} and nano-particle \cite{Petersen67} quantum emitters. More recently, focus has returned to the on-chip nano-photonic platform, using single QDs coupled to dielectric waveguides \cite{Coles2016,Coles2017,Mrowinski2019,Sollner2015,Barik666}. A notable strength of such an approach lies in harnessing the tightly-confined optical waveguide modes common to such a platform. This has the potential to enable highly efficient light-matter interactions at the single-photon level \cite{PhysRevLett.113.093603,PhysRevB.100.035311}, and is therefore of great interest for chiral quantum optics applications.

\begin{figure*}[ht!]
\centering
\includegraphics{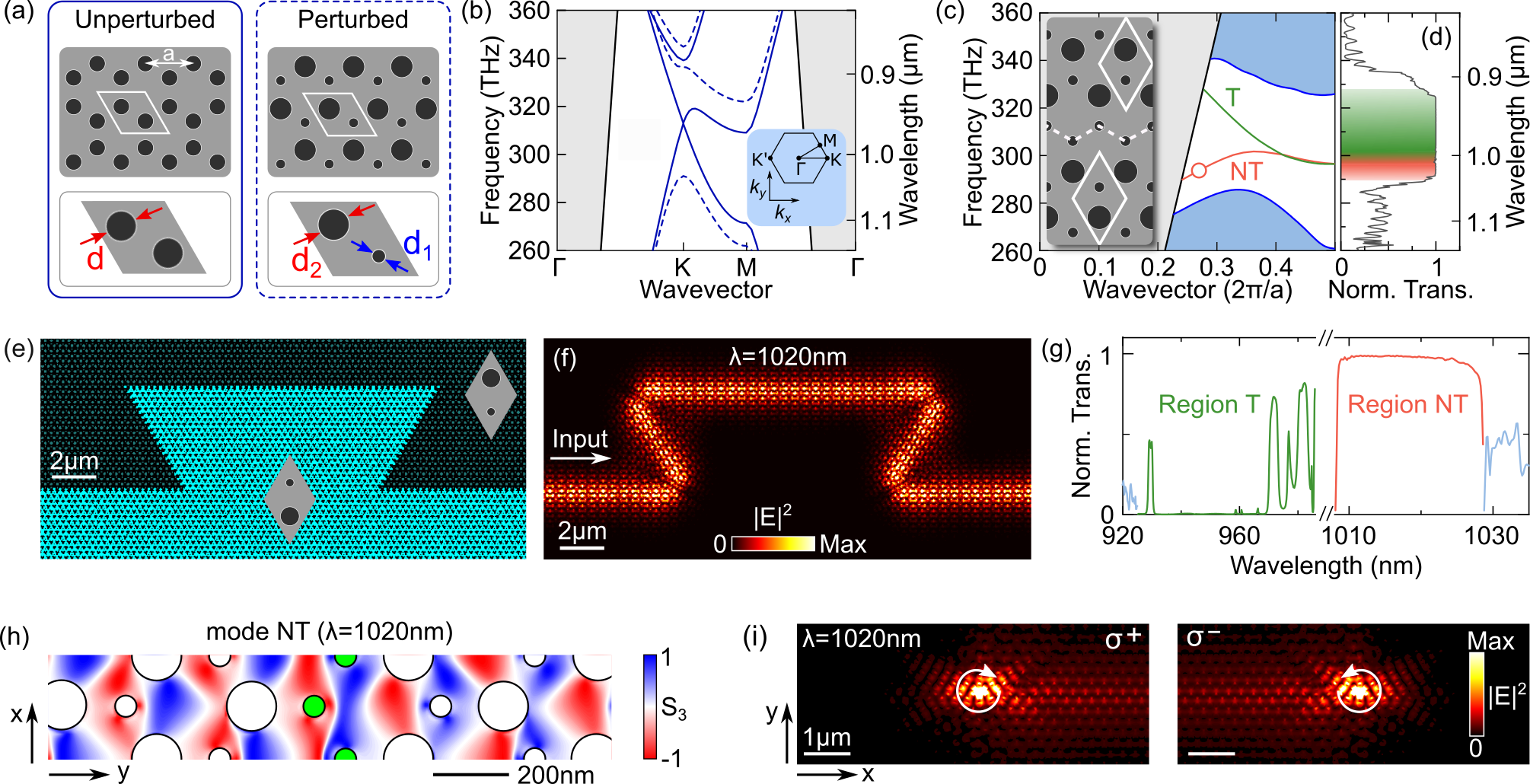}
\caption{(a) Schematic of a (left) unperturbed and (right) perturbed honeycomb lattice, formed from air holes in a thin dielectric membrane. The rhombic unit cell, comprising two holes, is outlined below in each case. (b) Band diagram for a honeycomb PhC formed using an unperturbed (solid lines) or perturbed (dashed lines) unit cell of the type shown in (a), in a 170nm-thick membrane (dielectric constant $\epsilon$=11.6). The inset shows the first Brillouin zone of the PhC.  Hole diameters are $d$=97nm, $d_1$=56nm and $d_2$=125nm. The pitch of the PhC in each case is $a$=266nm. (c) Projected band diagram for the interface between topologically-distinct perturbed photonic crystals (see schematic in inset, with the interface denoted by a dashed line). Holes sizes are the same as in (b). The interface supports two modes, labelled T (topologically trivial, see text) and NT (non-trivial). (d) FDTD-simulated normalised transmission coefficient (Norm. Trans.) for a 10$\mu$m length of the interface.(e) Schematic of a waveguide containing four 120-degree bends. Regions of opposite unit cell orientation are shaded light and dark, respectively. (f) Simulated time-averaged electric field intensity in the plane of the waveguide for mode NT ($\lambda=1020$nm), when light is injected from the left. (g) Normalised transmission through the waveguide in (f), colour-coded by spectral window (c.f. (c)). The abscissa break corresponds to the multimode region of the waveguide dispersion ($\sim986$nm-1008nm) which is not considered here. (h) Position-dependence of the normalised Stokes $S_3$ parameter at the interface, evaluated for mode NT at $\lambda$=1020nm (red circle in (c)). Circles represent the holes of the photonic crystal. Those with green fill show the position of the holes at the interface. Note that the waveguide is rotated 90 degrees compared to the schematic in (c). (i) Time-averaged electric field intensity resulting from a (left) $\sigma^{+}$ or (right) $\sigma^{-}$ circularly polarised dipole ($\lambda$=1020nm) placed at a chiral point of the waveguide corresponding to (left) $S_3\sim1$ or (right) $S_3\sim-1$.}
\label{fig:Interface-theory}
\end{figure*}

A concurrent development has seen the rise of topological photonics as a new paradigm in nano-photonics research \cite{PhysRevLett.100.013904,Hafezi2013,Rechtsman2013,RevModPhys.91.015006,Yasutomo2020}. Topological photonic interfaces are formed at the boundary between topologically-distinct photonic crystals and support the transport of light in counter-propagating waveguide modes which have helical character \cite{Wu2015}. This naturally suggests the possibility of a chiral topological photonic interface; the first such device was recently demonstrated using a QD coupled to a spin-Hall topological photonic crystal waveguide \cite{Barik666}. A particularly appealing property of topological waveguides is their predicted robustness against tight bends and certain defects \cite{PhysRevLett.100.013904}, which is attractive for the formation of low-loss, compact photonic elements. As an example, one could augment a chiral ring resonator \cite{chiral_resonator1,chiral_resonator2} with topological protection, an exciting prospect which would enable chiral coupling with enhanced light-matter interaction strength in a topologically protected system.

It is critical in nano-photonic design that waveguides restrict radiative coupling to free-space modes, which is a notable limitation of recently reported spin-Hall topological waveguides interfaced with QDs \cite{Barik666,MJ2019_APL,Spin-Hall_loss_theory}. This can be addressed by instead considering the valley-Hall topological photonic interface, for which the interface modes lie below the light line \cite{Shalaev2019,He2019}. Here, we realise a chiral quantum optical interface using semiconductor QDs embedded in a valley-Hall topological photonic crystal waveguide. Chiral coupling of single QDs to the non-trivial waveguide mode is demonstrated, with a spin-dependent, averaged directional contrast of up to $0.75\pm0.02$ measured. We investigate the propagation of light around tight bends in the topological interface by creating a compact ring resonator device. Q factors of up to 4,000 (125,000) are measured (simulated) for a resonator with a circumference of less than $17\mu$m. Finally, we couple the resonator to a bus waveguide, which enables us to demonstrate chiral coupling of a QD located within the resonator. In addition to a topologically non-trivial mode, the interface also supports a trivial mode, comparison with which enables a clear visualisation of the power of topological protection in the device.

\section*{Topological waveguide design}

Our valley-Hall topological photonic crystal (PhC) is formed from a honeycomb lattice of circular holes in a dielectric membrane, with the rhombic unit cell of the PhC comprising a pair of holes (see Fig.~\ref{fig:Interface-theory}a). Considering first the case of equivalent diameter holes, we plot in Fig.~\ref{fig:Interface-theory}b the band structure of the PhC for TE polarisation, revealing a Dirac cone at the $K$ point (and equivalently at the $K^\prime$ point, not shown). The band structure was calculated using the freely available MPB software package \cite{Johnson2001:mpb}. Next, we shrink one hole and expand the other, and show that the resulting PhC supports a bandgap for TE polarized light (dashed lines in Fig.~\ref{fig:Interface-theory}b). A key feature of the band structure is the opposite sign of the Berry curvature at the $K$ and $K^\prime$ points, as demonstrated in Ref \cite{He2019}. At an interface created by an inversion of the rhombic unit cells on one side of the PhC, the difference in Berry curvature leads to the confinement at the interface of counter-propagating edge states with opposing helicity \cite{He2019}. The band structure of such a waveguide, formed with a bearded interface, is shown in Fig.~\ref{fig:Interface-theory}c). The interface supports two modes, labelled T and NT, and is single mode between $\sim925$nm-986nm (region T) and  $\sim1008$nm-1028nm (region NT). In the multimode region in which the two modes overlap ($\sim986$nm-1008nm), the bands flatten and slow light is predicted~\cite{Endnote_1}. Using the approach of Ref. \cite{Yoshimi2020}, we gradually increase the size of the small holes on one side of the interface, transforming the bearded interface into a zigzag interface, consisting of two large holes facing one another. During this transformation we find that mode NT is always present in the band gap, whereas mode T disappears (see Supplementary Materials). We also observe in simulations that mode NT propagates smoothly around corners, whilst mode T is prone to back scatter (see Fig.~\ref{fig:Interface-theory}f-g and Fig.~\ref{fig:Resonator-theory}c-d). We therefore conclude that mode NT (T) is topologically non-trivial (trivial).

Using finite difference time domain (FDTD) simulations \cite{Lumerical} we show in Fig.~\ref{fig:Interface-theory}d that the normalised transmission for a 10$\mu$m-long section of the waveguide is approximately unity across the full spectral window covered by modes T and NT. A clear difference is observed, however, when waveguide bends are introduced. In Fig.~\ref{fig:Interface-theory}e-g we show the simulated transmission of light through a waveguide containing four 120-degree bends. Transmission of the topologically protected mode non-trivial mode is near unity within region NT. In contrast, transmission within region T is both significantly less than unity and wavelength dependent, due to backscatter of the trivial mode at each corner. This provides direct evidence of the topological protection afforded to mode NT.

Next, we demonstrate the potential of mode NT to form a chiral quantum optical interface. We evaluate the Stokes $S_3$ parameter (degree of circular polarisation) in the vicinity of the interface using FDTD simulations, revealing large areas in which $\left\lvert S_3\right\rvert\to1$ (see Fig.~\ref{fig:Interface-theory}h). Then, we position a circularly-polarised dipole source at a point of maximum chirality ($\left\lvert S_3\right\rvert\sim1$) and monitor the waveguide transmission (Fig.~\ref{fig:Interface-theory}i). Unidirectional emission with a direction dependent on the dipole polarisation is clearly predicted. Note that the same holds true for the trivial mode, in that case due to the breaking of mirror symmetry in the waveguide (see Supplementary Materials).

\section*{Experimental results}

\subsection*{Waveguide operation and chiral coupling}

\begin{figure}[t!]
\centering
\includegraphics[width=\linewidth]{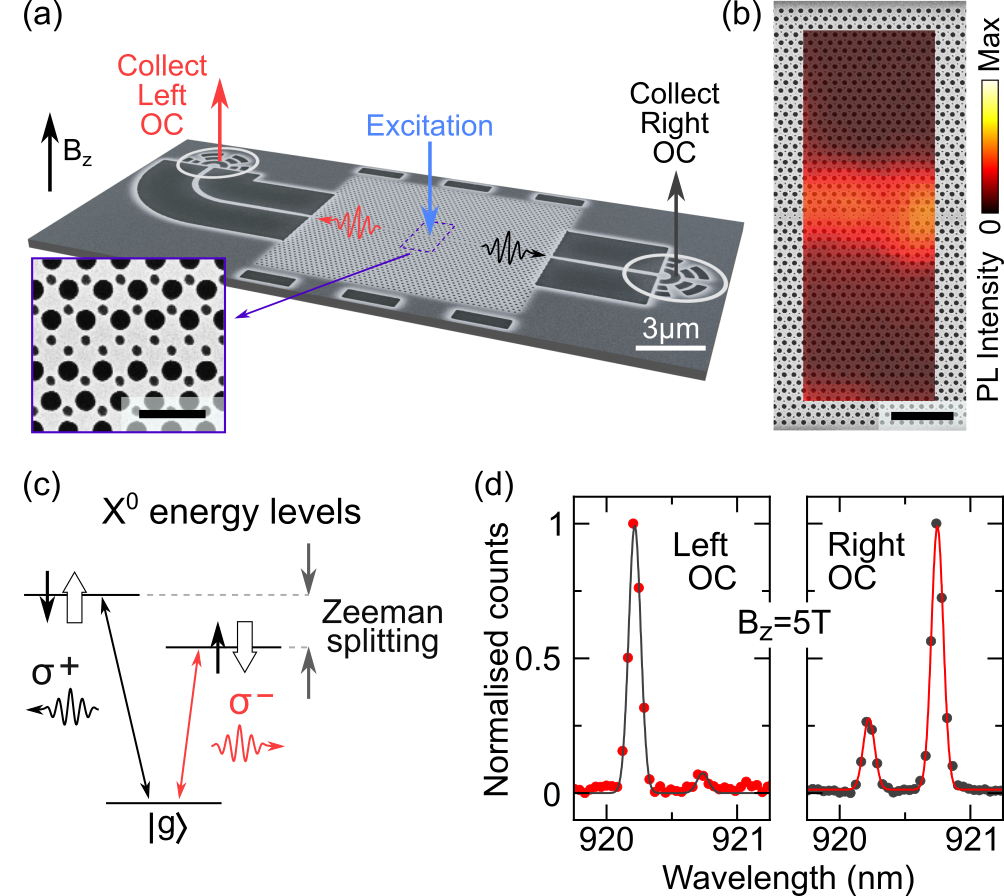}
\caption{(a) SEM image of a topological waveguide. A higher magnification image of the interface is shown in the inset (scale bar 500nm). (b) PL intensity collected from one outcoupler (OC) and integrated spectrally over region NT, as a function of the excitation position on the waveguide. The resulting map is overlaid on an SEM image of the device (scale bar $2\mu$m). The spatial resolution is limited by that of our optical microscope ($\sim$2$\mu$m). (c) Energy levels for the $X_0$ (neutral) exciton under non-zero magnetic field. (d) Normalised PL spectra at $B_z=5$T for a QD coupled chirally to mode NT, with a chiral contrast of $0.92\pm0.02$ ($0.57\pm0.03$) measured from the left (right) OC, respectively. Solid lines are the result of Gaussian fitting to the data (points). }
\label{fig:Interface-experiment}
\end{figure}

Topological PhC devices are fabricated in a nominally 170nm-thick GaAs \textit{p-i-n} membrane, which contains a layer of embedded InGaAs quantum dots (QDs). A scanning electron microscope (SEM) image of a representative topological waveguide is shown in Fig.~\ref{fig:Interface-experiment}a. The  waveguide is coupled at either end to nanobeam waveguides, which are terminated with grating outcouplers (OCs) to enable coupling of light to and from free space. Simulations show that the nanobeam - topological waveguide interface is efficient in region T ($>87\%$ transmission) but limited to an average of $31\%$ transmission in region NT (see Supplementary Materials), likely due to spatial mismatch between mode NT and the fundamental nanobeam TE mode. This is sufficient for optical readout of mode NT in the present experiment. It remains an outstanding challenge to create a highly efficient interface between trivial and topological waveguides, but encouraging progress has been reported elsewhere \cite{He2019}.

To demonstrate confinement of the optical mode at the topological interface, we collect PL from one OC while rastering the excitation laser across the device. After integrating the PL intensity spectrally over region NT, we show the signal as a function of excitation position in Fig.~\ref{fig:Interface-experiment}b, revealing that transmission of QD PL only occurs when the QD is in close proximity to the interface. Furthermore, we subsequently show (see Supplementary Materials) that QDs at the interface emit single photons and can be electrically Stark-tuned to control the wavelength of individual QD transitions. The latter is of particular interest for extending recent demonstrations of few-QD interactions in waveguide QED \cite{Kim2018,Grim2019} to the chiral regime supported by topological interfaces.

\begin{figure*}[t]
\centering
\includegraphics{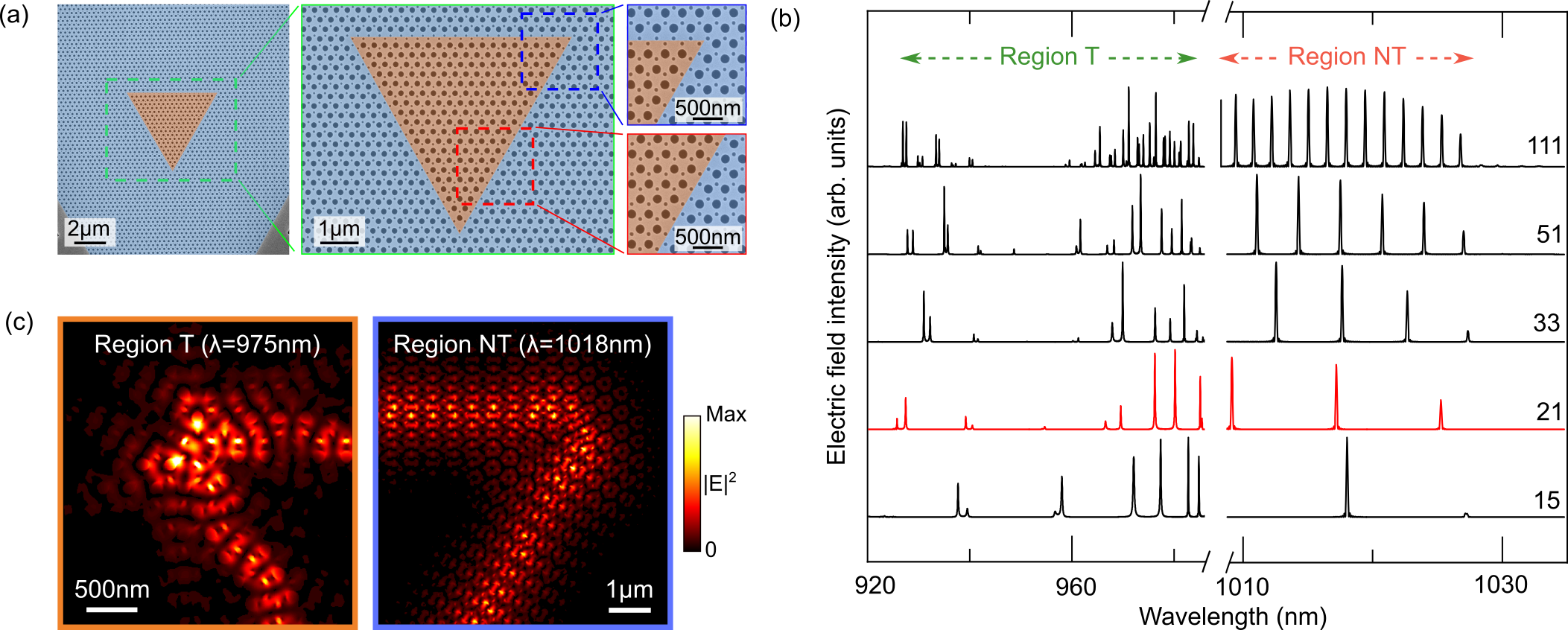}
\caption{(a) False-colour SEM images of a topological resonator, the centre of which is shaded orange. (b) Simulated longitudinal mode spectra as a function of resonator side length (spectra offset for clarity and labelled with the number of unit cells per side). The trace in red (21 unit cell side length) corresponds to the resonator size shown in (a). The abscissa break covers the multimode region of the waveguide dispersion. (c) Electric field intensity spatial profile in the corner of a resonator for (left) region T (975nm) and (right) region NT (1018nm).}
\label{fig:Resonator-theory}
\end{figure*}

Next, we probe the helicity intrinsic to the topologically non-trivial mode by investigating chiral coupling of QDs located at the interface. We first spectrally locate region NT of the waveguide under investigation (see Supplementary Materials). QDs emitting spectrally within this region are then optically excited in the presence of a magnetic field, which is applied in the Faraday geometry (normal to the sample plane). At the same time, PL is collected from both OCs. The magnetic field lifts the degeneracy of QD transitions via the Zeeman effect, allowing PL emission from states with opposite circular polarisation to be spectrally resolved (see Fig.~\ref{fig:Interface-experiment}c, for the case of a neutral exciton). The chiral contrast ($C$) is then evaluated independently for emission from either OC using the expression \cite{Coles2016}
$C=(I_{\sigma^+}-I_{\sigma^-})/(I_{\sigma^+}+I_{\sigma^-})$, where $I_{\sigma^+}$ and $I_{\sigma^-}$ refer to the PL intensity for $\sigma^+$ and $\sigma^-$ polarised emission, respectively. By considering each OC separately, any variance in collection efficiency is negated \cite{Sollner2015}. 

The resulting PL spectra for the emission from a single representative QD, for a magnetic field of B=5T, are shown in Fig.~\ref{fig:Interface-experiment}d. Two Zeeman-split states are observed, with asymmetric intensity for the $\sigma^+$ and $\sigma^-$ polarised transitions. The asymmetry in the intensities is seen to reverse when PL is collected from the other OC. The emission is therefore directional, with the intensity dependent on the handedness of the circular polarisation of the emitter (i.e. chiral coupling). The chiral contrast is as high as $0.92\pm0.02$ measured from the left OC. The contrast measured at the right OC is $0.57\pm0.03$, giving an average contrast of $0.75\pm0.02$. Asymmetry in the contrast measured in either direction is commonly observed in experiments such as this, an effect which as yet is not fully understood. Due to spatial symmetry breaking chiral coupling is also observed for QDs coupled to trivial mode T (see Supplementary Materials).

\subsection*{A chiral topological ring resonator}

\begin{figure}[t]
\centering
\includegraphics[width=\linewidth]{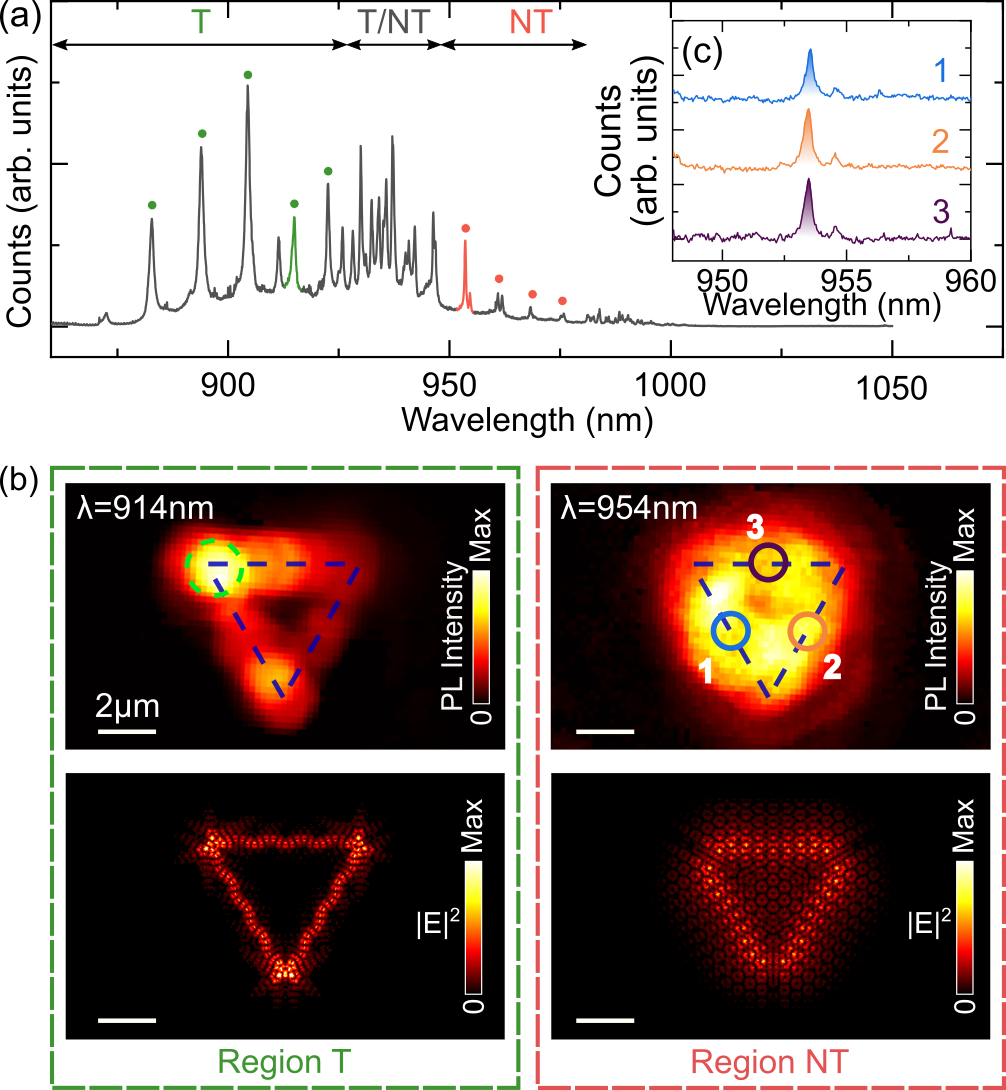}
\caption{(a) PL spectrum measured from above the resonator, with longitudinal modes visible (indicated in regions T and NT by filled circles). The different regions of the interface mode structure (T, T/NT and NT) are labelled. (b) (Upper panels) Spatially-resolved, integrated PL for two different modes of the same resonator, for wavelengths corresponding to regions T and NT, respectively (colour coded in (a)). PL was collected from the top left corner of the resonator (green dashed circle in the left hand panel). The excitation spot size was $\sim$2$\mu$m. (Lower panels) Simulated electric field intensity spatial profiles for modes in the same regions of the band structure considered in experiment. (c) PL spectra for a single longitudinal mode of the resonator in region NT, acquired at the midpoint of each of the three sides of the resonator, respectively (as labelled in (b)). The background has been removed from each of the spectra, which are offset for clarity.}
\label{fig:Resonator-experiment}
\end{figure}

\begin{figure}[b!]
\centering
\includegraphics[width=\linewidth]{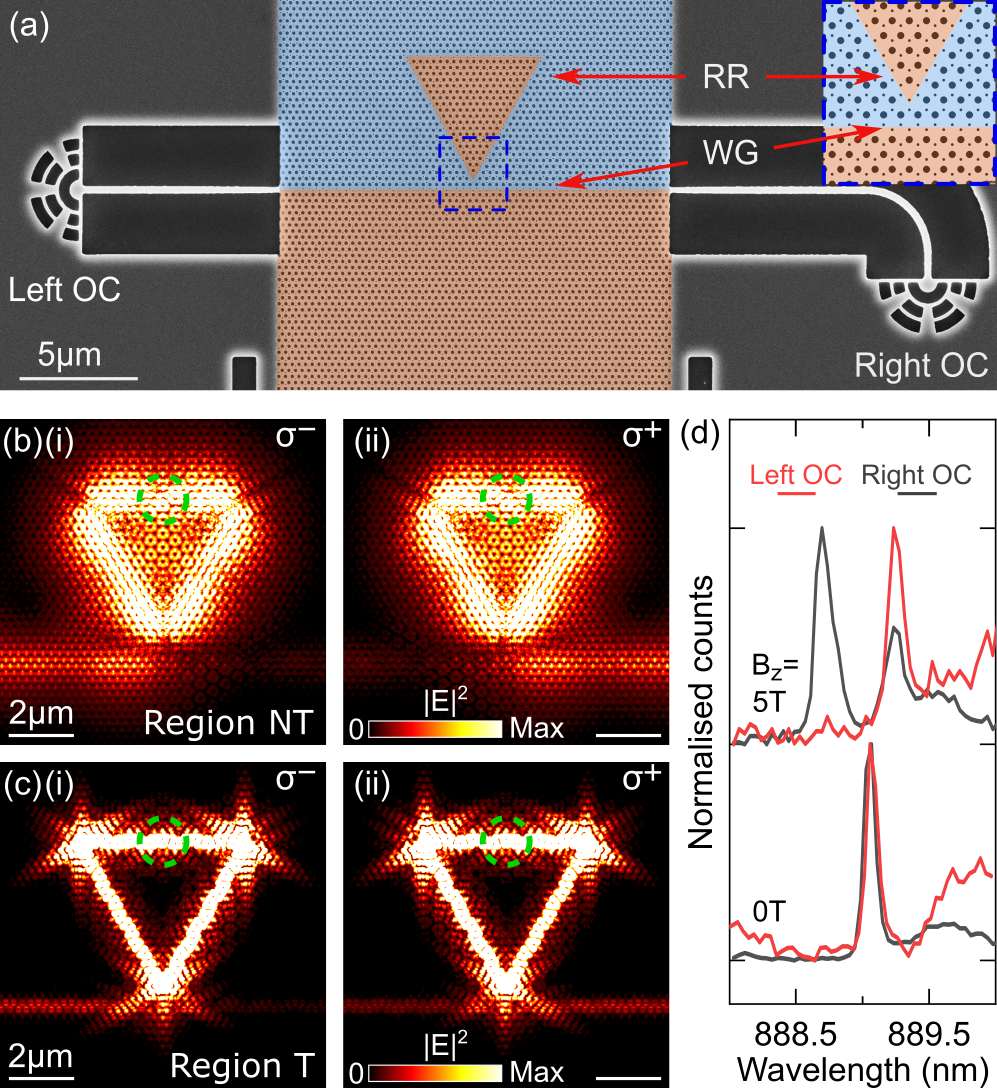}
\caption{(a) False-colour SEM image of a representative ring resonator (RR) coupled to a bus waveguide (WG), with nanobeam waveguides at either end. The inset shows a schematic of the RR-WG interface (dashed blue region in the SEM image). (b) FDTD simulated electric field intensity in the plane of the device for a (i) $\sigma^{-}$ and (ii) $\sigma^{+}$ polarised dipole source, located at a chiral point on the ring resonator interface (dashed green circle). The dipole emission wavelength corresponds to region NT of the waveguide dispersion. The colour scale is saturated to increase the visibility of the waveguide-coupled emission. (c) Same as in (b), but for an emission wavelength corresponding to mode T. (d) Low power, normalised PL spectra as a function of applied magnetic field, when the resonator is excited from above and PL is collected independently from both OCs. A single chirally-coupled transition can be seen, with chiral contrast of $0.93\pm0.07$ ($0.54\pm0.06$) from the left (right) OC. The emission wavelength corresponds to region NT. The shorter wavelength compared with elsewhere in the paper is due to larger hole sizes in this device.}
\label{fig:Coupled_wvg_RR}
\end{figure}

The combination of chiral coupling and an enhanced light-matter interaction strength is appealing for quantum optical device applications \cite{chiral_resonator1,chiral_resonator2,chiral_slow_light}. With this in mind, we now leverage the topological protection of mode NT against scattering at tight bends to create a compact photonic ring resonator, and demonstrate chiral coupling of a QD embedded within the device. The resonator takes the form of a triangle with three 120-degree corners (see Fig.~\ref{fig:Resonator-theory}a) and is created by embedding a triangular array of rhombic unit cells inside a larger array of inverted unit cells.

We first investigate the optical characteristics of the resonator using FDTD simulations. Longitudinal mode spectra as a function of the resonator side length are shown in Fig.~\ref{fig:Resonator-theory}b.  A clear distinction is seen in the resonator behaviour within the topologically non-trivial region NT and the trivial region T. In the latter, the longitudinal modes are split into multiple closely spaced resonances, which furthermore are strongly suppressed in the middle of the spectral range. These effects can both be understood as originating from back scattering of the trivial mode at the resonator corners. Simulation of a single 120-degree corner reveals a transmission minimum (i.e. strong back scatter) at $\sim950$nm, resulting in the suppression of the resonator modes (see Supplementary Materials). The simulated resonator mode quality factors (Q factors) for region T are limited by back scatter, with a maximum value less than 10,000 obtained for a device with a side length of $\sim5.5\mu$m (21 unit cells).

The spectrum corresponding to the non-trivial region NT stands in stark contrast. Here, the single corner transmission approaches unity across the full bandwidth. This finding is supported by a simulation of the mode profile in region NT, in which the corner is seen to be smoothly navigated, unlike the case in region T (see Fig.~\ref{fig:Resonator-theory}c). The mode spacing in region NT reduces with increasing resonator size, characteristic of a ring-type resonator. For the device with a side length of 21 unit cells the Q factor is as high as 125,000, an order of magnitude improvement on the trivial case. (Note that a simulated Q factor of greater than 1 million is obtained for a resonator with a side length of 111 unit cells.) This is direct evidence of the topological protection granted to mode NT.
 
Experimentally, we consider a resonator with a side length of 21 unit cells. We non-resonantly excite the ensemble of QDs on one side of the resonator at high power, creating a broadband internal light source, and collect PL emitted from the same location. A representative PL spectrum is shown in Fig.~\ref{fig:Resonator-experiment}a. The discrete modes of the resonator can be clearly resolved in both regions T and NT. The modes are observed at a shorter wavelength than that predicted by simulation, most likely due to an increase in fabricated hole sizes compared with design. Q factors up to $\sim$2,700 ($\sim$4,000) are measured in region T (NT). Higher Q factors in region NT compared to region T are consistent with topological protection of the former. The measured Q factors are considerably lower than those determined using simulations. This is commonly observed for GaAs devices operating in the 900-950nm wavelength range, and is likely due in part to surface-state-related absorption losses against which there is no topological protection. Surface passivation approaches have been shown to help mitigate this effect in GaAs photonic crystal devices \cite{Arakawa_160000_Q}.

We demonstrate confinement of the modes at the resonator interface by collecting PL emission from one corner of the resonator, while rastering the excitation laser across the device. At each excitation position, the PL spectrum is integrated over two different bandwidths corresponding to single longitudinal modes in regions T and NT, respectively. The resulting spatially-resolved PL maps are shown in Fig.~\ref{fig:Resonator-experiment}b, accompanied by simulated mode profiles. Correspondence between experiment and simulation is most easily seen by examining the corners of the resonator. In region T, strong scattering is predicted by simulation at the resonator corners, and the resulting experimental PL map reveals the distinct triangular shape of the full waveguide. In contrast, simulation shows that in region NT scattering is suppressed at the resonator corners (see Supplementary Materials). This effect can be seen clearly in the corresponding PL map, in which scattering at the resonator corners is very weak. To show that the mode spectra are consistent across the device, we plot in Fig.~\ref{fig:Resonator-experiment}c the PL spectrum acquired at the midpoint of each side of the resonator in region NT. The similarity in the spectra confirms that the mode is distributed along the interface.

Finally, we couple the topological resonator to an adjacent bus waveguide (see Fig.~\ref{fig:Coupled_wvg_RR}a). This enables us to demonstrate chiral coupling of a QD located within the resonator. (Similar coupling to a bus waveguide has been reported recently by Barik et al. \cite{Barik_2020}.) We first use FDTD simulations to demonstrate the principle of operation of the device. A circularly polarised dipole is positioned at a highly chiral point on the resonator interface, such that it emits unidirectionally. The optical field in the resonator subsequently couples evanescently to the bus waveguide. In the case of a wavelength within region NT, the direction of emission into the waveguide is seen to depend on the handedness of the dipole polarisation (Fig.~\ref{fig:Coupled_wvg_RR}b), therefore enabling readout of the chiral contrast. For the trivial mode T, this technique is hindered by back scatter at the resonator corners, and the maximum chiral contrast that can be measured is therefore reduced (Fig.~\ref{fig:Coupled_wvg_RR}c). We also find the chirality in this case to be sensitive to the position of the dipole along the interface (even when still positioned at one of the many highly chiral positions), unlike for the topologically protected mode NT. Experimentally, we excite QDs located at the resonator interface at low power, and measure the PL signal from the OCs at either end of the bus waveguide. Chiral coupling is shown in Fig.~\ref{fig:Coupled_wvg_RR}d for a QD transition at a wavelength corresponding to mode NT of the waveguide dispersion. The transition exhibits an average chiral contrast of $0.74\pm0.06$, demonstrating promising chiral characteristics for a topologically-protected ring resonator mode.

\section*{Conclusion}
We have demonstrated a chiral quantum optical interface using semiconductor QDs embedded in a valley-Hall topological waveguide. An average chiral contrast of up to 0.75 was measured for a QD coupled to a topologically non-trivial mode confined to the interface. Propagation of light around tight bends in the waveguide was subsequently demonstrated by fabrication of a topological photonic ring resonator, the modes of which have simulated Q factors up to 125,000 (experimental values up to 4,000). Coupling of the structure to a bus waveguide enabled the observation of chiral coupling of a QD within the resonator. Comparison with a topologically trivial mode in the same waveguide enabled us to clearly highlight the benefits of the topological protection which is afforded to the non-trivial mode. 

The topological nano-photonic platform demonstrated here has significant potential to form the basis of scalable chiral quantum optical circuits, protected against backscatter. Combining our approach with either QD registration \cite{Coles2016, Sapienza2015, Schnauber_QD_reg, Pregnolato_QD_reg} or site-controlled growth \cite{Reitzenstein_site_control,Jons_2013} techniques would enable deterministic positioning of QDs at highly chiral points in the waveguide, addressing the scalability challenge. Furthermore, the chiral nature of the interface allows for the realisation of separation-independent QD-QD interactions \cite{Lodahl2017,PhysRevB.92.155304}, in contrast to the nonchiral case. Exciting future prospects include the realisation of superradiant many-body states \cite{Grim2019,PhysRevB.96.115162} 
and the formation of large-scale chiral spin networks \cite{Pichler_2015} using a topologically-protected photonic platform.

\section*{Acknowledgments}

The authors would like to thank David Whittaker for valuable discussions.

\section*{Author Contributions}
M.J.M. designed the photonic structures, which were fabricated by R.D. E.C. and P.K.P. grew the sample. M.J.M. and A.P.F. carried out the measurements and simulations. L.R.W., M.S.S. and A.M.F. provided supervision and expertise. A.P.F. wrote the manuscript, with input from all authors. 

\section*{Funding Information}
Engineering and Physical Sciences Research Council (EPSRC) (EP/N031776/1).

\section*{Disclosures}

The authors declare no conflicts of interest.

\newpage
\bibliography{Main.bib}


\end{document}